\documentclass[aps,pre,twocolumn,groupedaddress,amsmath,amssymb]{revtex4-1}
\usepackage{ae}
\usepackage[pdftex]{graphicx}
\usepackage{amsmath}
\usepackage{amssymb}
\usepackage{epsfig, float}
\usepackage{hyperref}

\begin{document}

\date{\today }
\title{Individual heterogeneity generates explosive system network dynamics}


\author{Pedro D. Manrique and Neil F. Johnson}
\affiliation{Physics Department, University of Miami, Coral Gables, Florida FL 33126, U.S.A.}

\date{\today}

\begin{abstract}
Individual heterogeneity is a key characteristic of many real-world systems, from organisms to humans. However its role in determining the system's collective dynamics is typically not well understood. Here we study how individual heterogeneity impacts the system network dynamics by comparing linking mechanisms that favor similar or dissimilar individuals. We find that this heterogeneity-based evolution can drive explosive network behavior and dictates how a polarized population moves toward consensus. Our model shows good agreement with data from both biological and social science domains. We conclude that individual heterogeneity likely plays a key role in the collective development of real-world networks and communities, and cannot be ignored.
\end{abstract}


\maketitle

\section{\label{sec:level1}Introduction}
The mechanisms by which single entities (e.g. molecules, cells, people) come together as groups underlie myriad physical, biological and social processes \cite{Korniss,Soulier,Goncalves,palla,Estrada,Song,Caldarelli,barabasi,Fortunato}. Though multiple models have been proposed in order to explain the different aspects of real-world grouping behavior, the feature of heterogeneity among individuals has hardly been considered. Indeed, the universality of some collective processes across contrasting systems seems to have overshadowed the reality of intrinsically heterogeneous populations, and more importantly system-level evolution driven by individual heterogeneity. Percolation refers to the dynamical transition to large-scale connectivity by the addition of individual links \cite{Stauffer94,Sahimi94}. Before the transition, all clusters are of negligible size compared to that of the whole system. From the transition point onward, the largest cluster gathers a finite non-negligible fraction of the total number of individuals. The simplest way to form and grow clusters is by the classic Erd\'os-R\'enyi (ER) model \cite{ER59,ER60} which at each timestep a new link is established between two randomly selected nodes. This model reaches the percolation transition when the number of added links reaches half of the total number of nodes. At this point, the second moment associated with the cluster size distribution presents a singularity and the system passes to a gel phase \cite{Ziff82,Lushnikov06,redner10}.

\begin{figure}
\includegraphics[width=1\linewidth]{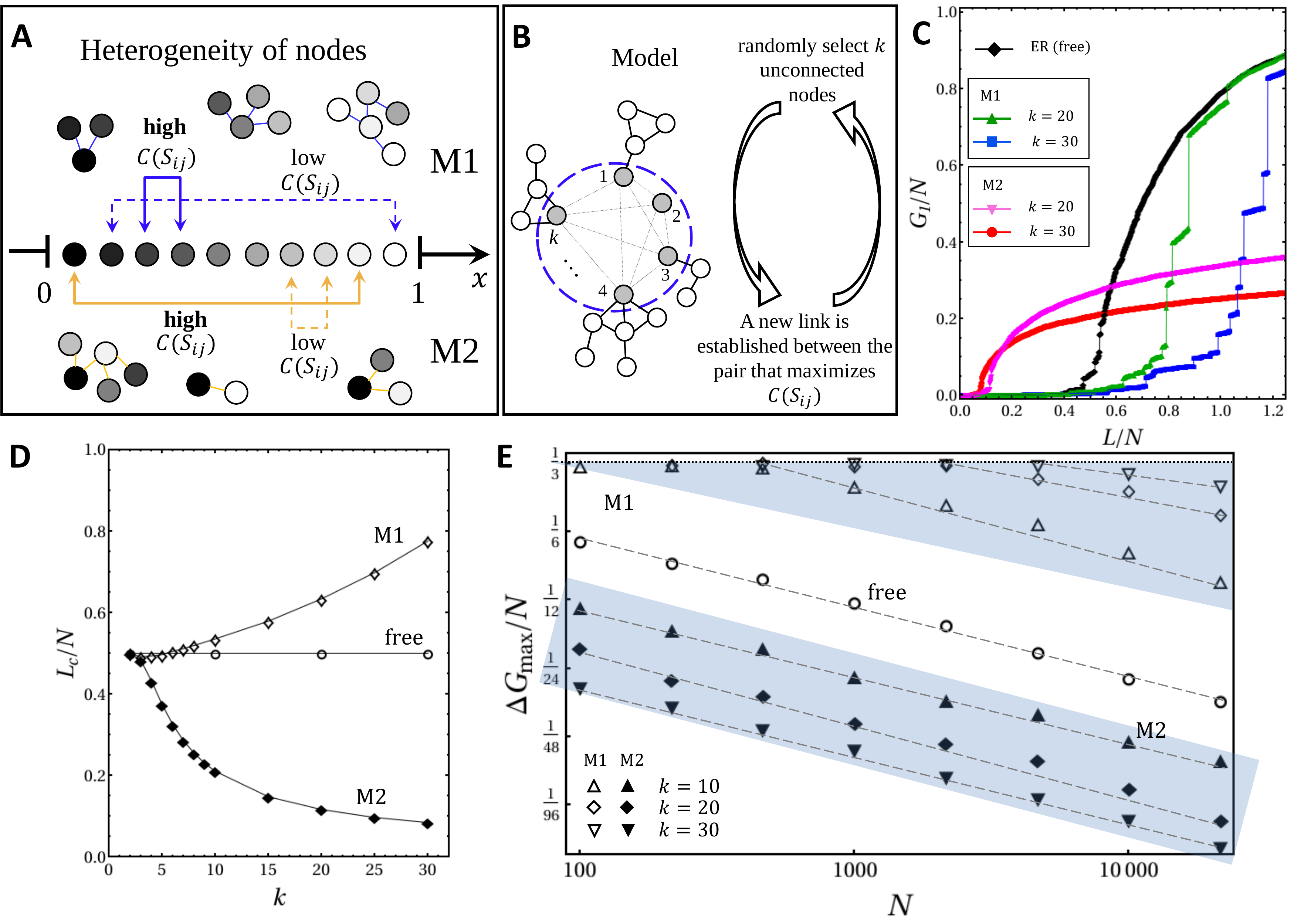}
\caption{\small{\textbf{Modeling heterogeneous grouping} A. The heterogeneity of nodes is modeled by the character $\{x\}$ which is a real number defined in the [0,1] interval and randomly assigned to each node from a distribution $q(x)$. Our model of interacting nodes comprises two mechanisms: M1 which favors linking similar nodes (i.e. close in character) and M2 favoring dissimilar nodes (i.e. distant in character). B. The model follows a two steps cycle where it first randomly selects a sample of $k$ unconnected nodes and second selects the pair that maximizes the coalescence function $\mathcal{C}(S_{i,j})$ that links together. C. Dynamics of the largest cluster $G_1$ as a function of the number of links for both heterogeneous formation mechanism as well as the random case (i.e. free) and different values of sampling size $k$. The total population is $N=10^4$ nodes. D. Variation of the transition point for M1, M2 and free model with the sampling size $k$. E. Dependence of the largest gap in the growth of $G_1$ for both mechanisms (shadows) and free model, and several values of sampling. Lines are guides to the eye. For D and E we show an average over $500$ realizations.}}
\end{figure}

Careful recent work has been dedicated to exploring the dynamical control of percolation, particularly the transition point \cite{Achlioptas09,Timme11,NPRev14}. Depending on the system, researchers look for ways to either delay or accelerate the appearance of the percolation transition. This is made possible by introducing the element of competition among the nodes to be associated together. Instead of randomly selecting one pair of nodes to be linked, three or more nodes are selected to compete for link addition \cite{Achlioptas09,Timme11}. The rules for addition vary from model to model, however the direct aspect that is being explored is how the different potential new links might affect the size of the resultant new cluster. Thus, rules have been proposed so that only small clusters are formed, which delays the transition, or large clusters join resulting in the contrary effect. These new models of large-scale connectivity show abrupt variations in the size of the largest cluster and hence are known as {\em explosive} percolation. Examples from the biological and social domain have been shown to have features akin to explosive percolation \cite{Donati06,Hernan10,Bounova12}. However, it is not obvious why these models should be able to ignore individuals' heterogeneity when determining which clusters are formed and when. Even in our everyday lives, we do not arbitrarily form clusters without some underlying factor determined by the characteristics of the individuals involved -- for example in social gatherings, we often join with family members who by definition have similar genes to us, while in a sport we join those having complementary skills in order to form a strong team. Our findings suggest that heterogeneity-based cluster formation and hence node-to-node affinity interactions can play a crucial role in generating explosive phenomena in real-world systems, and influence the point of transition. As a result, our work provides fresh insights into how the diversity of individuals could affect the overall dynamics of the system. Moreover, the output of our models closely captures the dynamics of real-world systems such as online extremism and protein homology networks. 

\section{Model}
The heterogeneity is introduced as a hidden variable $x_i$ that we call the `character' and is assigned randomly to each node $i$ from a distribution $q(x)$ \cite{Char1,Char2}. For simplicity we consider all $\{x\}$ to be real numbers between $0$ and $1$ since larger ranges can be easily rescaled to be within this interval. Also, we consider that the character values are constant in time though this can be modified to include variations over time to account for experience or external influence. The mechanisms of link addition follow directly from the relationship among the $\{x\}$ values associated to the nodes to be linked. This is quantified by the similarity $S_{ij}$ between node $i$ and node $j$ which is defined as $S_{ij}=1-|x_i-x_j|$. Thus highly similar nodes are close to each other in the $x$ spectra and otherwise for highly dissimilar nodes. The mechanisms of link addition rest in the definition of the coalescence function $\mathcal{C}(S_{ij})$. We consider two complementary mechanisms for link addition: mechanism 1 (M1) favoring similar nodes and mechanism 2 (M2) favoring dissimilar nodes (see Fig.1A). A system following M1 tends to form groups of alike individuals (e.g. kin) while M2 tends to form groups with unlike or complementary individuals (e.g. teams).  Hence, for M1 a coalescence function is defined as $\mathcal{C}(S_{ij})=S_{ij}$ while for M2, $\mathcal{C}(S_{ij})=1-S_{ij}$.

Our model starts with all $N$ nodes unconnected. At each timestep a sample from the system is randomly selected and a new link is established between the pair of nodes that maximizes the coalescence function $\mathcal{C}(S_{ij})$ (see Fig. 1B). Thus, the distribution $q(x)$, the sampling method and size, and the specific mechanism of link addition (i.e. M1 or M2), dictate the evolution of the system. Though the sampling can be either of nodes or links, the evolution of the network present similar properties. Sampling of nodes can be related to individuals that ran into each other in the course of a time period (e.g. a single day) and among all the interactions only a few connections are established based on their mutual affinity. Here we will consider sampling of nodes for simplicity, with the link version in the Supplemental Material (SM). The smallest sampling size refers to two randomly selected nodes which sets the limit of a random graph model (ER model). Note that this is independent of the link addition mechanism and the distribution $q(x)$. In a similar way considering a distribution $q(x)$ to be a Dirac Delta, implies that all the nodes are identical (i.e., character free) and hence the limit of a random graph model is also found independently of the remaining parameters. The former and latter observations talk about the competitive and diverse aspects of the network formation process, respectively. Unless both aspects are present the system will follow an ER process. 

The evolution of the system is typically described by means of the size of the largest cluster that we call $G_1$. All present clusters sizes are denoted $G_{i}$, where $i$ is the size rank starting from the largest (i.e., $i=1$). Figure 1C shows the evolution of $G_1$ for several sampling sizes $k$, and mechanisms M1 and M2 when the character values are assigned randomly from a uniform distribution $q(x)$. The aggregation mechanisms and sampling sizes have contrasting macroscopic effects in the evolution of $G_1$. For example, for M1 the percolation transition shifts from the random case (character free) point at $L/N=0.5$ to greater values of $L/N$ and exhibits macroscopic features akin to explosive percolation such as large jumps. On the other hand, for M2 the transition shifts to smaller values of $L/N$ than the random case and the growth tends to be smoother. In all cases the shift tends to be more severe as the sampling size increases as shown in Fig. 1D. 

These features can be explained by means of the formation mechanisms. M1 links pairs of nodes whose $x$ values are the closest to each other. The probability density function (PDF) $f(y)$ of the similarity $y=S_{i,j}$ associated to the uniform distribution is $f(y)=2y$ for $y\in [0,1]$. Since the distribution is maximum at $y=1$, there is a large number of pairs with high similarity that could lie at any point in the character spectra. Hence, the formation of small and medium size clusters in all $x$ regions is expected. This explains the appearance of jumps in the evolution of the largest cluster resulting from the aggregation of small and medium-size clusters to $G_1$. On the other hand, M2 tends to link nodes that lie far from each other in the character spectra. The PDF tells us that the number of optimal pairs for M2 process (i.e. small or zero $S_{ij}$ and hence $y$) is low. As a consequence, for large samples it is very likely that either of these optimal nodes already belong to the largest component and therefore the formation of medium size clusters is less probable. Thus, the growth of $G_1$ for M2 becomes gradual and smoother than that of M1.

Figure 1E illustrates this fact by examining the size of the largest gap $\Delta G_{max}$ in the evolution of $G_1$ ($\Delta G_{max}=\text{max}\{G_1(L+1)-G_{1}(L)\}$) due to the addition of a single link. It clearly shows that the gaps associated to M2 are far smaller than those of M1. Interestingly we find that the gap scales algebraically with the system size as shown in other explosive percolation models \cite{Timme11,NPRev14} and this behavior is independent of the link addition mechanism. However, it is found that there is an upper bound in the largest gap for M1 around $1/3$ when the sampling size reaches a specific fraction of the whole system. This observation can be understood as follows. For a uniform character distribution the mean field character difference (i.e. $\langle|x_{i}-x_{j}|\rangle$) between any two nodes is $1/3$. This means that, for a large sampling size, the optimal pair would carry a character value difference smaller than or around $1/3$. Thus, linking nodes that are separated for more than $1/3$ in character becomes highly unlikely. This process gives rise to the formation of two or three large clusters at separated points of the character spectra that fuse when they have expanded enough and since the distribution is uniform the growth in size follows the expansion in character and hence the gap is on average $1/3$.

\begin{figure}
\includegraphics[width=1\linewidth]{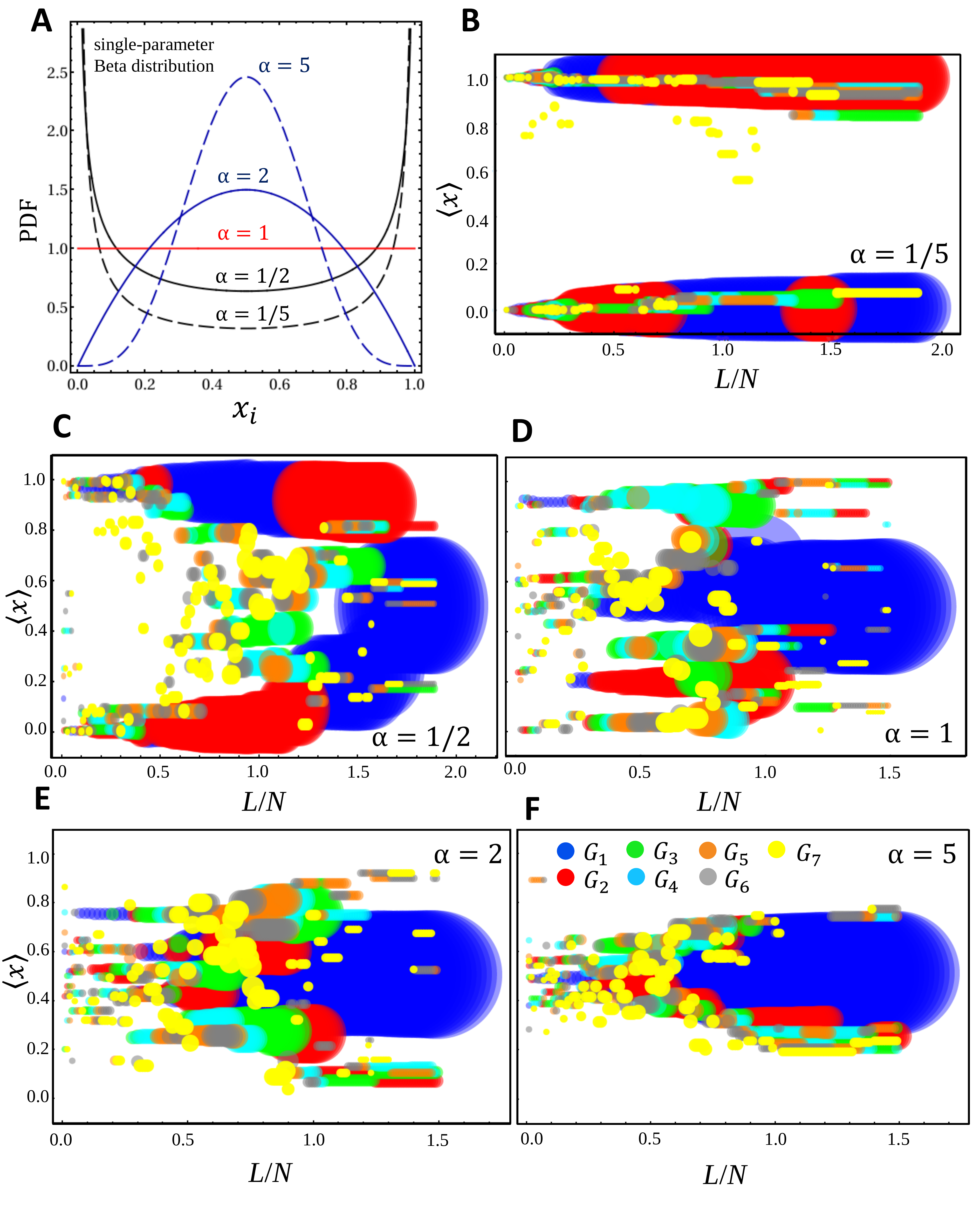}
\caption{\small{\textbf{Effect of diversity in grouping}. A. Probability density function of Beta distribution for several values of the parameter $\alpha$ for the case when $\alpha=\beta$. B-F. Grouping evolution as a function of the average character of the largest seven clusters (colored bubbles). The size of the bubble is proportional to the square root of the size and each panel shows a different $\alpha$ value as shown in each panel. The total population is $N=10^3$ nodes, sampling $k=10$ and M1 aggregation rule. }}
\end{figure}

Next, we analyze the diverse aspect which refers to the character distribution $q(x)$. To this end, here we consider $q(x)$ to be equal to a single parameter Beta distribution ($\beta=\alpha$) which is symmetric around $x=0.5$ as shown in Fig. 2A. This choice allows us to look at polarized populations but with the same average character value. The uniform distribution previously presented is found for the special case of $\alpha= 1$. Polarized populations can be represented by $0\leq\alpha<1$, where the severity in the polarization increases as alpha approaches zero. In the limit of $\alpha = 0$, the system is maximally polarized with half of the nodes having character equal to $1$ and half equal to $0$. This binary system follows two independent random graph formation processes that will join together only after both graphs are fully formed (see SM). By contrast, unpolarized populations result for $\alpha>1$ where the limit of $\alpha \rightarrow \infty$ corresponds to the random graph process is found since the character values of all nodes are identical.

\begin{figure}
\includegraphics[width=1\linewidth]{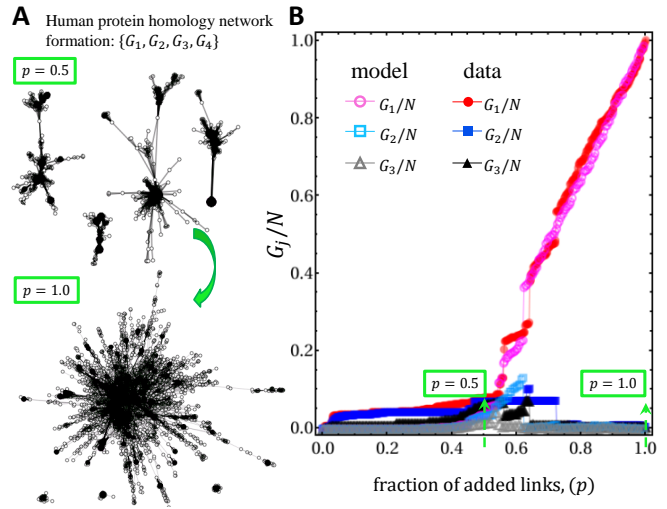}
\caption{\small{\textbf{Heterogeneity in the Human Protein Homology Network} A. Snapshots of the human protein homology network as links are added. The parameter $p$ represents the fraction of added links. Top panel is the early stage for $p=0.5$ and bottom panel is the final stage with $p=1$. Each panel shows the largest four clusters (i.e. $G_{i}$, for $i=1,2,3,4$). B. Evolution of the three largest clusters of the protein network showed in A (filled symbols) compared to those of a heterogeneous process (empty symbols) with a uniform character distribution and a sampling size of $k=8$.}}
\end{figure}

Figures 2(B-F) illustrate single simulation results of the evolution of the seven largest clusters (colored bubbles) vertically positioned at their average character value when the population follows a single parameter Beta distribution with different $\alpha$ values. The size of the bubble is proportional to the square root of the number of nodes within the cluster and the colors represent the rank from one to seven according to their size (see legend at Fig. 2F). Here we look at the grouping dynamics for the case of M1 while M2 will be presented in the SM. For $\alpha<1$, the distribution is highly polarized and groups are formed at opposite extremes of the character spectra. Depending on the severity of the polarization the system would require more or less links to reach consensus around $x=0.5$. This is because as $\alpha\rightarrow 0$, the number of nodes at the center of the $x$-spectra (e.g. moderates) is smaller and links between groups at opposite poles becomes less likely (see Fig. 2B). A less severe polarization (e.g. $\alpha=1/2$) allows for a certain number of moderates which serve as bridges between the groups in the poles and help reach consensus (see Fig. 2C). Populations with comparable number of nodes along each portion of the $x$ spectra (i.e. $\alpha=1$) form small and medium size clusters at different $x$ points that subsequently join together into the largest component around $x=0.5$. Interestingly, after consensus is attained, some extreme groups tend to appear at both poles (see Fig. 2D). This behavior is also present for a distribution with a larger amount of moderates than extremists as depicted in Fig. 2E for the case of $\alpha=2$. Finally, for $\alpha=5$ we find that the extremes are formed around the edges of the consensus group as shown in Fig. 2F. These results can provide some insights concerning how to create consensus (e.g. political) in a diverse and even polarized population of interacting individuals, but warn of the possibility of leaving residual isolated pockets of individuals with rather extreme average values of $x$ (i.e. away from $0.5$). 

\section{Real-world networks}
We now explore two different real network systems from the biological and social science domains that experience explosive grouping behavior.

\subsection{Protein Homology Network}

Networks of proteins can be generated by identifying homology relationships, i.e., commonalities in the amino acid sequences of a pair \cite{Donati06,Bates06,SIMAP}, and connecting them accordingly. Here proteins are viewed as nodes that are linked to each other through weighted edges. In principle, the network can be comprised of all deduced proteins. For simplicity, here we look into the subset of human proteins which have shown features akin to explosive percolation \cite{Hernan10}. The link's weight is determined by the homology relationship of a given pair. Highly homologous proteins have a greater weight than heterogeneous proteins.
The homology between a given pair and hence the weight of their connecting link is measured by the alignment score $s_{ij}$ while the score accuracy is determined by the Expectation ($E$) value \cite{Lipman90,Pearson00}. The smaller the $E$-value the more reliable the score $s_{ij}$ becomes \cite{Lipman90,Pearson00}. According to UniProt \cite{UniProtKB}, a total of $159,522$ human proteins are deduced which can be divided in $20,214$ that have been reviewed manually against $139,338$ that await revision. Here we analyze the scores among the subset of reviewed human proteins provided by the Similarity Matrix of Proteins project (SIMAP) \cite{SIMAP} and we have used links with $E$-values up to $10^{-10}$. For simplicity, we use the score ratio (SR) as a weight measure which is defined as: $SR=s_{ij}/\max{\{s_{ii},s_{jj}\}}$, where $s_{ii}$ is the self-homology score. Note that $SR$ is defined within the $[0,1]$ interval where $1$ indicates perfect alignment. The system starts with all proteins isolated and links are added from high to low according to their alignment. Thus highly homologous communities are formed first and links between different communities are established later.

Figure 3 presents our findings for the protein system. Fig. 3A illustrates two snapshots of the evolution of the four largest protein clusters (i.e. $G_{i}$, for $i=1,2,3,4$) for a mid-point evolution stage (top panel) where half of the links have been added, and the final stage (bottom panel) where the last link has been added illustrating their contrasting topologies. Fig. 3B shows that the size of the largest cluster (red circles) tends to show explosive dynamics as new links are added. This behavior is captured by our heterogeneous grouping model (pink rings) using a uniform distribution $q(x)$, M1 formation and sampling size of $k=8$. Both systems experience explosive grouping behavior with comparable rates and gaps. Interestingly, our model also simultaneously reproduces some of the features in the dynamics of the second and third largest clusters (squares and triangles, respectively). The agreement tends to be higher after the percolation transition than earlier. We attribute this to the strong homology in some protein communities where single nodes can act as hubs gathering many individuals and rising to a non-negligible size. In our model, all nodes are equally likely to be sampled for potential addition. Moreover, the restrictions inflicted by the $E$-value leave many potential links absent which explains why at the last stage the network is not fully unified (see Fig.3A). Despite these complications in the alignment parameters, our model is still in reasonable greement with several features of the evolution of the protein network. This is an indication of the wide flexibility that a heterogeneity framework brings to the grouping behavior which makes it adaptable to different dynamical systems.


\subsection{Online grouping}
We next consider the online social group formation in support of Islamic State (ISIS) whose data was collected in Ref. \cite{Science16}. This occurs on Europe's largest social media platform based in Russia, VKontakte (VK, https://vk.com). As of December 2017, this site counts with 460 millions of users worldwide and has been used by the extremists to spread propaganda and to recruit sympathizers. A snapshot of the pro-ISIS network is presented in Fig. 4A for January 10, 2015 where $59$ different extreme groups where active with a total of $21,881$ followers combined establishing $48,605$ connections (i.e. follows). This platform has become ideal for extreme groups in part because similar networking services such as Facebook shut down these type of online groups almost immediately, while VK takes more time to act. During that active time the online groups attract followers and grow in particular ways. The methodology of the data extraction is presented in Ref \cite{Science16} where different evolutionary adaptations have also been uncovered. For example, groups can change names, restart a new group after a shut down or switch between visible and invisible preferences in order to avoid moderators and being shut down, among other. 

\begin{figure}
\includegraphics[width=1\linewidth]{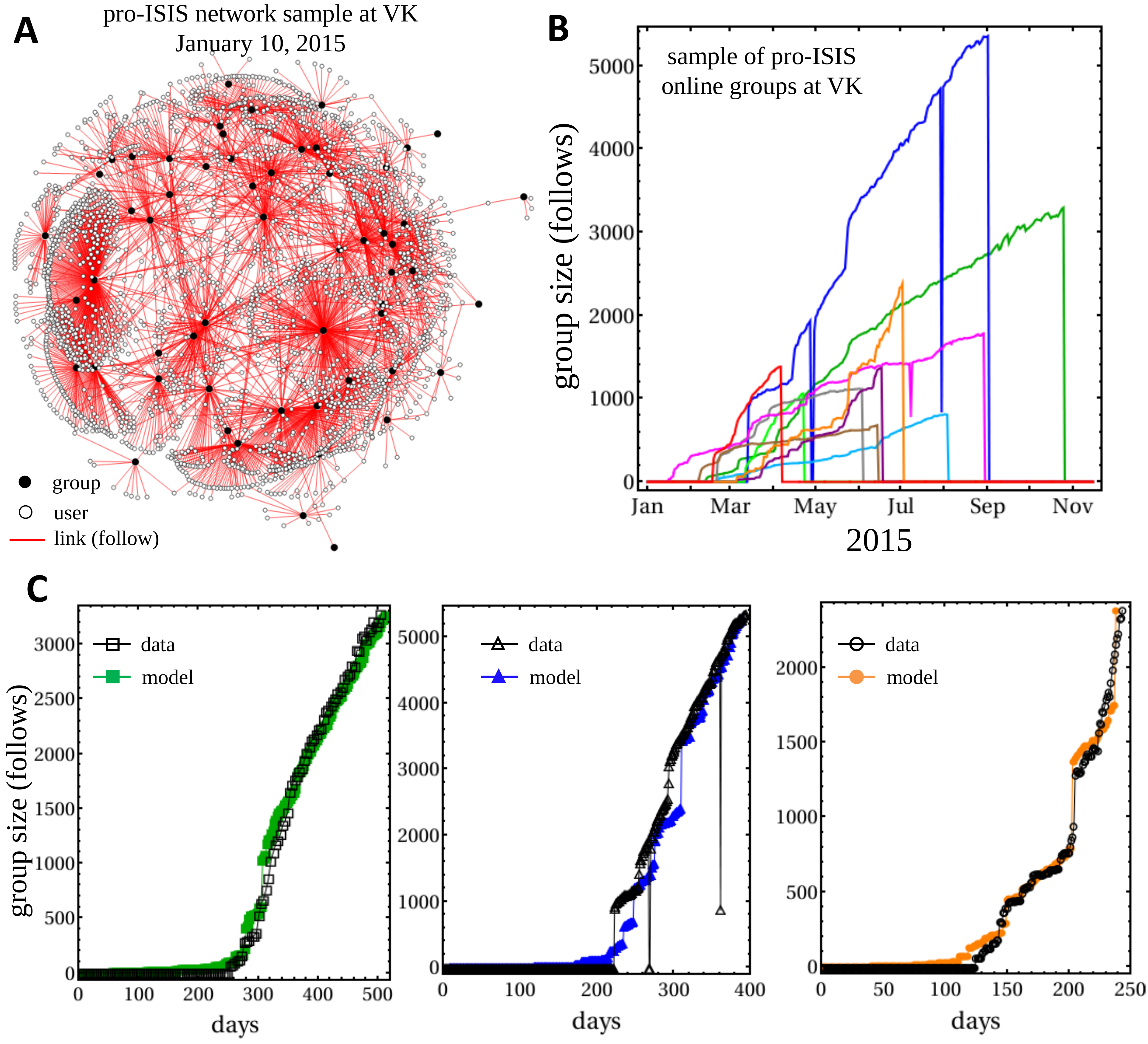}
\caption{\small{\textbf{Explosive grouping in pro-ISIS online groups} A. Snapshot of online pro-ISIS support groups on the VK platform on January 10, 2015. B. Evolution of a sample of sharkfin pro-ISIS groups from first detection to shutdown. C. Examples of explosive behavior in pro-ISIS groups as compared to our heterogeneous model for $k=5$ (left panel) and $k=10$ (middle and right panel). For each case $N$ is set to be the respective group population at the moment of shutdown.}}
\end{figure}

Figure 4B shows the evolution of a sample of extreme online groups from the time of their earliest detection up to the moment where they were shut down, at daily resolution. The size of the groups is determined by the number of users that decide to follow them. We note that in this particular sample the size of the group passes from zero to an average size of $200$ follows with a maximum of nearly $1000$ follows in a single day. In addition, we see that this irregular growth is repeated at several sections of the formation process. These irregular jumps in the size at the start or during the evolution of a given group $g_1$, could be the consequence of a group (or groups) $g_0$ being shut down and hence all its (or their) former members coordinate to either join group $g_1$ likely due to affinity with their message and generating a jump in its size, or to open a new group $g_2$ and hence creating a jump at the start of its evolution. Note that the latter process indicates that group $g_2$ is a continuation of group $g_0$ and therefore they are essentially the same group, while the former is a cluster aggregation process. In both cases the changes in the sizes are abrupt and an association with a random aggregation process is less accurate. We therefore propose that a heterogeneous percolation model with group formation M1 cannot be discarded as a potential mechanism for the creation and subsequent growth of these particular groups. 

Figure 4C strengthens our proposal by capturing key features of the formation of extreme online groups. The panels show how our heterogeneous model compares with three of the extreme groups shown in Fig. 4B (color indicates the specific modeled group). The remaining groups are compared with the model in Fig. S6 of the SM. The model interprets the VK system as a collection of several sub-systems, each with a specific sub-population of potential follows that aggregate over time and whose largest component grows to become the extreme group. Note that the sub-population is not of users but `follows' since each user can follow several groups simultaneously. Potential followers explore groups daily and decide whether to join or not, arguably based on affinity. This can be considered a competitive process where only some users (either isolated or from former shut down groups) add to the extreme group's population. Hence we implement a competitive modeling where on each timestep $k$ nodes compete for addition. Our results of Fig. 4C as well as Fig. S6 show that values of $k$ between $2$ and $10$ capture general growth trends as well as some key features such as the size jumps. Due to the bipartite nature of the group evolution, the formation process considers inter-link additions only. This framework allows us to estimate the start of the online activity even when the group was invisible or not yet sufficiently extreme, and hence did not appear in the data collection radar. Also, this modeling opens the gate for exploration of different intervention strategies to mitigate the spreading of the group by attacking it, for example, at its earliest stage. We note that not all the groups identified can be modeled by this explosive percolation framework for technical reasons, e.g. because of missing data.

\section{Summary}
We have shown that a heterogeneous population of interacting individuals can generate explosive grouping behavior. In addition, our model provides a framework to study the impacts of new links on polarized populations. Linking individuals can result in the formation of new residual clusters at the extremes. We also tested our model against two different heterogeneous real-world datasets capturing specific features of the formation process and showing that heterogeneity plays a decisive part in the system's network evolution.

\section{Acknowledgments}
We thank Chaoming Song for discussions regarding the model, Minzhang Zheng and Yulia Vorobyeva for assistance with the pro-ISIS data and Thomas Rattei for providing the protein data. N. F. J. is grateful to the National Science Foundation (NSF) grant CNS1522693 and Air Force (AFOSR) grant FA9550-16-1-0247. The views and conclusions contained herein are solely those of the authors and do not represent official policies or endorsements by any of the entities named in this paper.


\begin{thebibliography}{99}

\bibitem{Korniss} M. Anghel, Z. Toroczkai, K.E. Bassler and G. Korniss, Phys. Rev. Lett. {\bf 92}, 058701 (2004)
\bibitem{Soulier} A. Soulier and T. Halpin-Healy, Phys. Rev. Lett. {\bf 90}, 258103 (2003)
\bibitem{Goncalves} B. Goncalves and N. Perra, {\em Social Phenomena: Data Analytics and Modeling} (Springer, Berlin, 2015) 
\bibitem{palla} G. Palla, A.L. Barabasi, and T. Vicsek, Nature {\bf 446}, 664 (2007)
\bibitem{Estrada} E. Estrada, Phys. Rev. E {\bf 88}, 042811 (2013)
\bibitem{Song} C. Song, S. Havlin and H. Makse, Nature Phys. {\bf 2}, 275 (2006)
\bibitem{Caldarelli} G. Caldarelli, {\em Scale-Free Networks: Complex Webs in Nature and Technology} (Oxford University Press, Oxford, 2007)
\bibitem{barabasi} A.L. Barabasi and H.E. Stanley, \textit{Fractal Concepts in Surface Growth} (Cambridge University Press, 1995)
\bibitem{Fortunato} F. Radicchi and S. Fortunato, Phys. Rev. E {\bf 81}, 036110 (2010)
\bibitem{Stauffer94} Stauffer, D. \& Aharony, A. {\em Introduction to Percolation Theory} (Taylor \& Francis, 1994) 
\bibitem{Sahimi94} Sahimi, M. {\em Applications of Percolation Theory} (Taylor \& Francis, 1994)
\bibitem{ER59} Erd\"os, P. \& R\'enyi, A. On random graphs I. Math. {\em Debrecen} {\bf 6}, 290-297 (1959) 
\bibitem{ER60} Erd\"os, P. \& R\'enyi, A. On the evolution of random graphs {\em Publ. Math. Inst. Hungar. Acad. Sci.} {\bf 5}, 17-61 (1960).
\bibitem{Ziff82} E.M. Hendriks, M.H. Ernst, and R.M. Ziff. Coagulation equations with gelation. {\em J. Stat. Phys.} {\bf 31}, 3, (1983)
\bibitem{Lushnikov06} A. A. Lushnikov. Gelation in coagulating systems. {\em Physica D}, {\bf 222}, 37-53, (2006).
\bibitem{redner10} P.L. Krapivsky, S. Redner and E. Ben-Naim, \textit{A Kinetic View of Statistical Physics} (Cambridge University Press, Cambridge, 2010)
\bibitem{NPRev14} D'Souza R. M. \& Nagler, J. Anomalous critical and supercritical phenomena in explosion percolation. {\em Nature Physics} {\bf 11}, 531-538 (2014).
\bibitem{Timme11} Nagler, J., Levina, A. \& Timme M. Impact of single links in competitive percolation. {\em Nature Physics} {\bf 7}, 265-270 (2011)
\bibitem{Achlioptas09} Achlioptas, D., D'Souza, R. M. \& Spencer, J. Explosive percolation in random networks. {\em Science} {\bf 323}, 5920, 1453-1455 (2009).
\bibitem{Hernan10} Rozenfeld, H. D., Gallos, L. K. \& Makse, H. A. Explosive percolation in the human protein homology network. {\em Eur. Phys. J. B} {\bf 75} 305-310 (2010).

\bibitem{Bounova12} G. Bounova \& O. de Weck. Overview of metrics and their correlation patterns for multiple-metric topology analysis on heterogeneous graph ensembles. {\em Phys. Rev. E.} {\bf 85}, 016117, (2012)
\bibitem{Donati06} D. Medini, A. Covacci, \& C. Donati. Protein homology network families reveal step-wise diversification of type III and type IV secretion systems. {\em PLoS Computational Biology} {\bf 2}, 12, e173, 1543-1551 (2006).
\bibitem{Char1} N. F. Johnson, P. Manrique and P. M. Hui. {\em J. Stat. Phys.} {\bf 151}, 395 (2013)
\bibitem{Char2} P. D. Manrique, P. M. Hui and N. F. Johnson. {\em Phys. Rev. E} {\bf 92}, 062803 (2015)
\bibitem{Bates06} P.F. Jonsson, T. Cavanna, D. Zicha \& P.A. Bates. Cluster analysis of networks generated through homology: automatic identification of important protein communities involved in cancer metastasis. {\em BMC Bioinformatics}, {\bf 7}, 2 (2006)
\bibitem{SIMAP} Thomas Rattei, Roland Arnold, Patrick Tischler, Dominik Lindner,Volker St\''umpflen \& H. Werner Mewes. SIMAP: the similarity matrix of proteins, {\em Nucleic Acids Research}, {\bf 34}, D252-D256 (2006).
\bibitem{Lipman90} S.F. Altschul, W. Gish, W. Miller, G. Myers, and D.J. Lipman. A basic local alignment search tool. {\em J. Mol. Biol.}, {\bf 215}, 403-410 (1990)
\bibitem{Pearson00} W.R. Pearson. Flexible sequence similarity searching with the FASTA3 program package. {\em Methods Mol. Biol.}, {\bf 132}, 185-219 (2000)

\bibitem{UniProtKB} The UniProt Consortium. UniProt: the universal protein knowledgebase. {\em Nucleic Acids Res.} {\bf 45} D158-D169 (2017)

\bibitem{Science16} Johnson, N. F., Zheng, M., Vorobyeva, Y., Gabriel, A., Qi, H., Velasquez, N., Manrique, P., Johnson, D., Restrepo, E., Song, C. \& Wuchty, S. New online ecology of adversarial aggregates: ISIS and beyond. {\em Science} {\bf 352}, 6292, 1459-1463 (2016).

\end{thebibliography}

\end{document}